\documentclass[a4paper,11pt]{article}
\usepackage{soul}
\usepackage{jinstpub} 

\usepackage[mathlines]{lineno}
\usepackage{tikz}

\title{\boldmath  Methods for the study of light propagation in LArTPCs}

\author{Adames, M. R.}
\affiliation{Universidade Tecnológica Federal do Paraná,\\
3165 Avenida Sete de Setembro, Brazil}

\emailAdd{marcioadames@utfpr.edu.br}

\abstract{
Liquid Argon Time Projection Chambers (LArTPCs) are widely used in particle physics experiments. They use light and charge released in events to reconstruct and analyze them. Light information collected by the Photon Detection System (PDS) is used mainly to determine the initial time stamp, $t_0$, and to support determining the total energy of the event. For these purposes, it is important to simulate and predict the propagation of light inside the detector. There are several approaches to this, and this work discusses some of the options.
}


\arxivnumber{2501.19363} 

\begin{document}
\maketitle
\flushbottom

\section{Introduction}

Simulations of light propagation through diverse media are important across a diverse array of scientific disciplines, ranging from astrophysics and particle physics to medical imaging. The characterization of photon density, flux, radiance, and related concepts can be simulated using different methods, which present advantages and disadvantages in different contexts.

These methods include Monte Carlo simulations that consider the behavior of each individual photon, which can be developed in Geant4, for example. They take full physics considerations into account and can be very accurate. Geant4 is a software toolkit that simulates particles passing through matter \cite{Geant4}. This method might not be the most adequate for some applications because it can be very intense computationally and demands the whole system to be simulated, even if the interest might be in localized detectors inside that system. Another approach used by researchers is to try to predict light propagation using differential equations.

A classical approach based on differential equations is to use the Radiative Transfer Equation (RTE), which is a non-linear integro-differential equation that takes into account the most relevant aspects of photon propagation, such as absorption and scattering behaviors through matter. Depending on the assumptions of Transport Theory, this might lead to high-dimensional differential equations \cite{Transport_Theory}, which might be very hard to solve analytically and be cumbersome to solve numerically. We do not give an account of this vast subject in this work but limit ourselves to mentioning that there are many solutions to specific cases and numerical methods proposed in the literature.  

Often, for practical applications, simplifying assumptions might be made on the RTE to make it easier and faster to solve. Such an approximation for photon propagation is the Photon Diffusion Equation, which is also vastly studied in the literature, e.g., \cite{Martelli}. The diffusion approximation can be solved analytically in many cases but might not be adequate for some applications because it allows for infinite propagation speed (does not respect causality) and does not account for ballistic-like photon propagation for short distances. Boundary effects might also be cumbersome to consider in this setting.

Other models have been proposed, such as relativistic diffusion equations, to mitigate the inadequacies of the diffusion equation. Lemieux, Vera, and Durian propose a hyperbolic partial differential equation for photon propagation in scattering and absorbing media in \cite{Lemieux:1}. This work considers their equation in a large cuboid media (in opposition to the thin slabs often investigated for propagation in living tissue) and presents an analytical solution for a pulse-like source in unbound media adapted to the LArTPC context.

New approaches to the problem of light propagation prediction are still appearing in the literature, and this work also describes shortly the semi-analytical approach of  \cite{Szelc}, which was developed in the context of particle physics.

\section{Geant4 simulations (Monte Carlo)}

Monte Carlo simulations are widely used to model light propagation in complex media, particularly in applications such as medical imaging, radiotherapy, and particle physics. Geant4 is a powerful software toolkit that enables the simulation of the passage of particles and radiation through matter, including light (photons). It provides a flexible and accurate framework for simulating various processes such as absorption, scattering, and reflection, which are essential for understanding light-matter interactions.

The core principle of the Monte Carlo method in this context is the random sampling of physical processes and trajectories to model the behavior of photons as they interact with the medium. The Geant4 simulation framework handles the following steps:

\begin{enumerate}
\item Photon Generation: Photons are emitted from a source with a defined energy spectrum, direction, and position.
\item Interaction Tracking: As photons propagate through the medium, they interact with the material via processes such as absorption, scattering, and reflectance. Geant4 provides detailed models for these interactions based on the material properties.
\item Event Handling: Each photon’s journey is tracked until it either escapes the system or gets absorbed.
\item Detector Response: Simulation can capture the final interactions of the photon, such as where it is detected or how it contributes to the measured signal.
\end{enumerate}

The Geant4 toolkit provides a comprehensive set of classes for photon tracking and interaction processes. It allows for flexible configuration and optimization, making it suitable for contexts ranging from high-energy physics to medical applications. By performing simulations for a large number of emitted photons, the Monte Carlo method provides statistical insights into how light interacts with materials and can be used to simulate and optimize experimental setups. These simulations can be computationally intensive if the system/detector is large or has an intricate geometry, which might cause photons to go through many interactions with the medium (absorptions, re-emissions, scatterings, reflections, \dots).

\section{Radiative Transfer Equation}

The RTE considers that, as a beam of radiation travels, it loses energy to absorption and redistributes energy by scattering. There are different formulations to it, depending on the coordinate system and contexts of the situation. Following the formulation of \cite{machida2024}, it is an equation on the specific intensity, $I(\mathbf{r}, \hat{\mathbf{s}}, t)$, at position $\mathbf{r}$ in direction $\hat{\mathbf{s}}$ at time $t$
$$
\frac{1}{v}\frac{\partial}{\partial t}I(\mathbf{r}, \hat{\mathbf{s}}, t) + \hat{\mathbf{s}} \cdot \nabla I(\mathbf{r}, \hat{\mathbf{s}}, t) + (\mu_{a} +\mu_s) I(\mathbf{r}, \hat{\mathbf{s}}, t) = \mu_s \int_{\mathcal{S}^2} p(\hat{\mathbf{s}},\hat{\mathbf{s}}') I(\mathbf{r}, \hat{\mathbf{s}}, t)  d\hat{\mathbf{s}'} + S(\mathbf{r}, \hat{\mathbf{s}}, t)
$$
where $v$ is the speed of light in the medium, $\mu_s$ is the scattering coefficient, $\mu_a$ is the absorption coefficient, $p(\hat{\mathbf{s}},\hat{\mathbf{s}}')$ is the scattering phase function and $S(\mathbf{r}, \hat{\mathbf{s}}, t)$ is a source term.

This equation accounts for the intensity of light in every direction at any point and time, and thus is a high-dimensional integro-differential equation, which can be hard to solve analytically, so it is often solved numerically in practical applications.

\section{Photon Diffusion Equation}

Some situations are modeled by the simple diffusion equation, which might not be accurate enough for many applications. So we focus this work on the Photon Diffusion Equation (PDE), which describes the rate of change in spherical irradiance in a scattering and absorbing medium. It is commonly used to model light transport in situations where multiple scattering events occur, and it assumes that the photons are diffusing in the medium. The PDE written for the spherical irradiance (or fluence rate), $\Phi (\mathbf{x},t)$, is:
$$
\frac{\partial}{v\partial t}\Phi(\mathbf{x},t) - D \Delta \Phi(\mathbf{x},t) + \mu_a \Phi(\mathbf{x},t) = \delta(t) \delta(\|\mathbf{x}-\mathbf{x}_0\|),
$$
where $\delta$ is Dirac's delta function, $\Delta = \nabla^2$ is the Laplacian in the spatial coordinates, as formulated by \cite{Martelli}. In this formulation, the diffusion constant, $D$, is given in terms of the absorption and scattering coefficients, the anisotropy constant, $g$, of the medium and a constant $\gamma$, with different references using different values for it, usually $0\leq \gamma \leq 1$:
$$
D = \frac{1}{3(\gamma \mu_{a} + (1-g)\mu_{s})}.
$$
The reference \cite{Martelli} makes a more detailed discussion of the value of $\gamma$. 

The source term \(  \delta(t) \delta(\|\mathbf{x}-\mathbf{x}_0\|) \) represents the injection of photons into the medium, often associated with a laser or point like light pulse.

This equation can be solved analytically or numerically and the solution provides insights into the propagation of light, including how the intensity or fluence of photons decreases as they diffuse through the scattering media. This equation is often used in biological tissue to model near-infrared spectroscopy and imaging techniques.

Absorbing / reflective walls can be considered through the subtraction/addition of mirror image solutions to the equation, with sources outside of the detector, as presented by \cite{Galymov}.

This equation fails to describe some characteristics of light because it does not capture the ballistic-like behavior of photons for short distances, it allows for faster-than-light photons and has difficulty in considering boundary effects since making the photon density zero at the boundary also makes the flux zero at it. To solve this last issue, researchers have considered the zero boundary actually outside the physical wall, which causes further problems in some contexts.

\section{Relativistic diffusion}

The proposal of \cite{Lemieux:1} takes the exact form of the transport equation in one-dimensional space and assumes that the three-dimensional equation is identical in form. They obtained a Telegrapher's equation for the photon density $\varphi$:
\begin{equation}
\label{TelEq} \Delta \varphi = \frac{\partial^2 \varphi}{v^2 \partial t^2} + \left(\frac{2}{\lambda_{abs}}+\frac{3}{\lambda_{rs}^*}\right)\frac{\partial \varphi}{v \partial t} + \frac{1}{\lambda_{abs}}\left(\frac{1}{\lambda_{abs}}+\frac{3}{\lambda_{rs}^*}\right)\varphi,
\end{equation}
where $v$ is the speed of light in the medium, $\lambda_{abs}$ is the absorption length, $\lambda_{rs}^* = \lambda_{rs}/(1-g)$ is an increased absorption length and accounts for the anisotropy through dividing the scattering length, $\lambda_{rs}$, by 1 minus the anisotropy constant $g$.

The chosen numerical coefficients in equation \ref{TelEq} guarantee:
\begin{itemize}
    \item The correct ballistic behavior, so that the equation should approach a damped wave equation with speed $v$ as $\lambda^*_{rs} \to \infty$.
    \item Diffusive limits, so that the equation should approach the relativistic heat conduction with speed $v$ as $\lambda_{abs} \to \infty$.
\end{itemize}
For further clarification, we refer to \cite{Lemieux:1}.

For liquid argon we are assuming, based on experimental data \cite{babicz}, \cite{jones} and \cite{Galymov}:
\begin{align}
    v &= 21.7 cm/ns\\
    \lambda_{abs} &= 2000cm,\\
    \lambda_{rs} & = 100cm,\\
    g & =0.0025.
\end{align}

An analytic solution to this equation for a pulse point source in an unbound medium can be calculated explicitly. Tautz and Lerche calculated an analytical solution to equation
\begin{equation}
\frac{\partial f}{\partial t} + \tau \frac{\partial^2 f}{\partial t^2} = \kappa_{||} \frac{\partial^2 f}{\partial z^2} + \kappa_{\perp} \left(\frac{\partial^2 f}{\partial x^2} + \frac{\partial^2 f}{\partial y^2}\right) \label{tautz_eq}
\end{equation}
in \cite{equat}, where $\tau, \kappa_{\perp}$ and $\kappa_{||}$ are constants. Adapting their solution to our coordinate system and units and looking for solutions to eq. \ref{TelEq} of the form $\varphi (\mathbf{x},t) = f (\mathbf{x},t) e^{-\alpha t}$  it is possible to derive a solution to eq. \ref{TelEq}:
\begin{align*}
\varphi(t, \mathbf{x}) =& \frac{3^{3/2} e^{-\frac{v}{2}\left(\frac{2}{\lambda_{abs}}+\frac{3}{\lambda_{rs}^*}\right) t}}{4\pi (v\lambda_{rs}^*)^{3/2}}\left[\frac{\delta(t - \|\mathbf{x}-\mathbf{x}_0\|/v)}{\|\mathbf{x}-\mathbf{x}_0\|}\sqrt{\frac{v\lambda_{rs}^*}{3}} I_0\left(\frac{3v}{2\lambda_{rs}^*}\sqrt{t^2 - \frac{\|\mathbf{x}-\mathbf{x}_0\|^2}{v^2}}\right)\right.\\
& +\left.\frac{\sqrt{3v}}{2\sqrt{\lambda_{rs}^*}} \frac{H\left(\sqrt{3v/\lambda_{rs}^*} (t - \|\mathbf{x}-\mathbf{x}_0\|/v )\right)}{\sqrt{t^2 - \|\mathbf{x}-\mathbf{x}_0\|^2/v^2}}  I_1\left(\frac{3v}{2\lambda_{rs}^*}\sqrt{t^2 - \frac{\|\mathbf{x}-\mathbf{x}_0\|^2}{v^2}}\right)\right]\\
 = &  \varphi_{wave}(t, \mathbf{x}) + \varphi_{dif}(t, \mathbf{x}),
\end{align*}
where  $I_0$ and $I_1$ are modified Bessel functions of the first kind, $H$ is the Heaviside function and $\delta$ is Dirac's delta function.

This solution is composed of the sum of a pulse-like wave moving with speed $v$:
\begin{equation}
    \label{photon_den_wave}\varphi_{wave}(t, \mathbf{x}) = \frac{3^{3/2} e^{-\frac{v}{2}\left(\frac{2}{\lambda_{abs}}+\frac{3}{\lambda_{rs}^*}\right) t}}{4\pi (v\lambda_{rs}^*)^{3/2}}\frac{\delta(t - \|\mathbf{x}-\mathbf{x}_0\|/v)}{\|\mathbf{x}-\mathbf{x}_0\|}\sqrt{\frac{v\lambda_{rs}^*}{3}} I_0\left(\frac{3v}{2\lambda_{rs}^*}\sqrt{t^2 - \frac{\|\mathbf{x}-\mathbf{x}_0\|^2}{v^2}}\right),
\end{equation}
and a diffusive component contained inside the sphere defined by the pulse-like component:
\begin{align}
    \label{photon_den_diff} \varphi_{dif}(t, \mathbf{x}) = \frac{3^{3/2} e^{-\frac{v}{2}\left(\frac{2}{\lambda_{abs}}+\frac{3}{\lambda_{rs}^*}\right) t}}{4\pi (v\lambda_{rs}^*)^{3/2}}\frac{\sqrt{3v}}{2\sqrt{\lambda_{rs}^*}} \frac{H\left(\sqrt{3v/\lambda_{rs}^*} (t - \|\mathbf{x}-\mathbf{x}_0\|/v )\right)}{\sqrt{t^2 - \|\mathbf{x}-\mathbf{x}_0\|^2/v^2}}  I_1\left(\frac{3v}{2\lambda_{rs}^*}\sqrt{t^2 - \frac{\|\mathbf{x}-\mathbf{x}_0\|^2}{v^2}}\right).
\end{align}

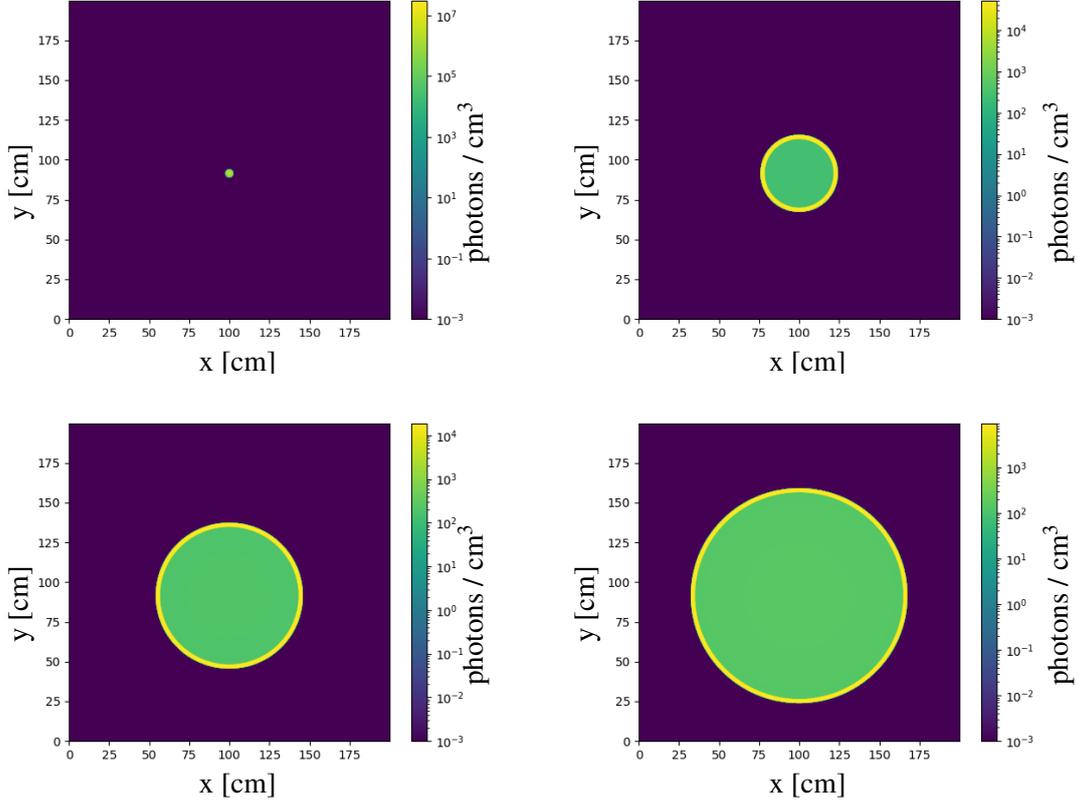
\begin{figure}[ht!]
\centering
\begin{tikzpicture}
    \draw (-1.3, 0) node[inner sep=0] {\includegraphics[trim = 50 23 0 0, clip, width=65mm]{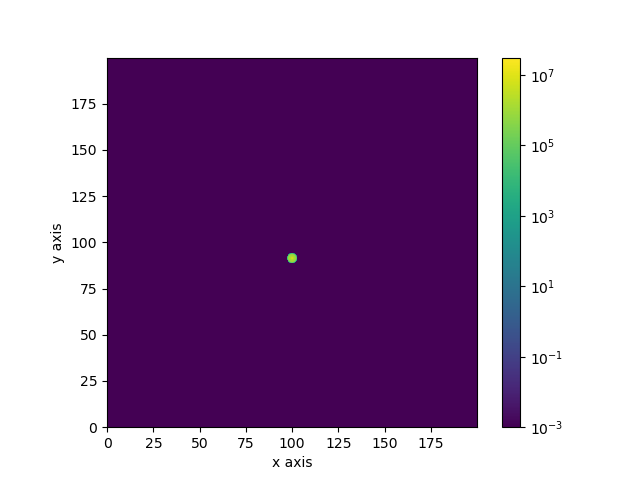}};
    \draw (-1.9, -2.9) node {x [cm]};
    \draw (-4.75, -0.5) node {\rotatebox{90}{y [cm]}};
    \draw (1.2, -0.5) node {\rotatebox{90}{photons / cm$^3$}};
    \draw (6.2, 0) node[inner sep=0] {\includegraphics[trim = 50 23 0 0, clip, width=65mm]{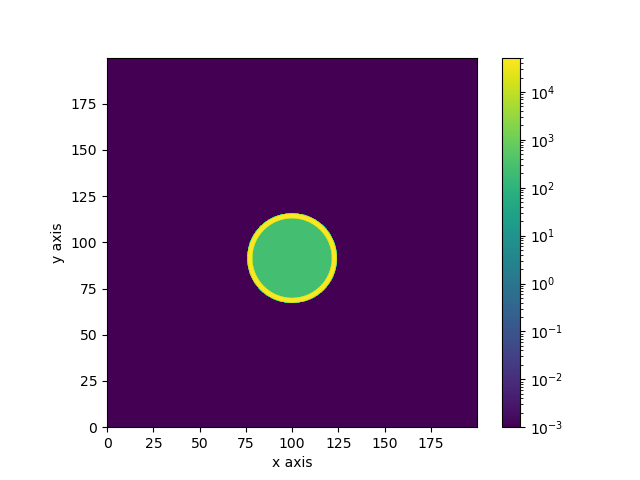}};
    \draw (5.6, -2.9) node {x [cm]};
    \draw (2.7, -0.5) node {\rotatebox{90}{y [cm]}};
    \draw (8.9, -0.5) node {\rotatebox{90}{photons / cm$^3$}};
    \draw (-1.3, -5.6) node[inner sep=0] {\includegraphics[trim = 50 23 0 0, clip, width=65mm]{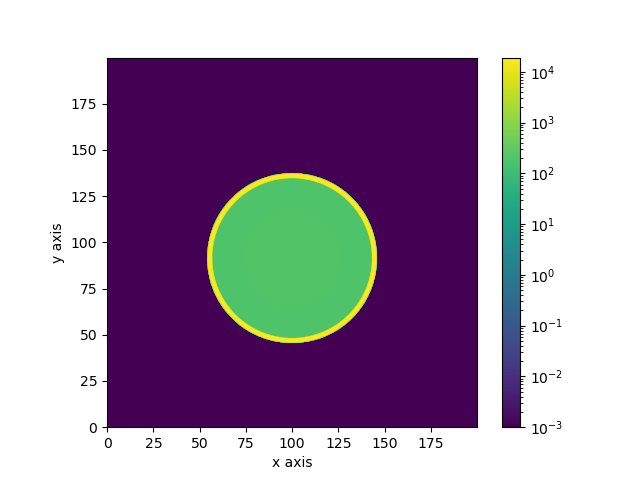}};
    \draw (-1.9, -8.5) node {x [cm]};
    \draw (-4.75, -6.1) node {\rotatebox{90}{y [cm]}};
    \draw (1.2, -6.1) node {\rotatebox{90}{photons / cm$^3$}};
    \draw (6.2, -5.6) node[inner sep=0] {\includegraphics[trim = 50 23 0 0, clip, width=65mm]{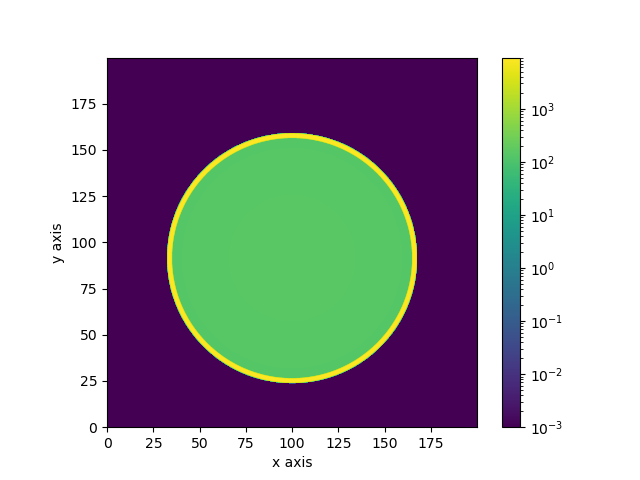}};
    \draw (5.6, -8.5) node {x [cm]};
    \draw (2.7, -6.1) node {\rotatebox{90}{y [cm]}};
    \draw (8.9, -6.1) node {\rotatebox{90}{photons / cm$^3$}};
\end{tikzpicture}
    \vspace{-5mm}\caption{The different figures illustrate the photon density in the square $ \{0 \leq x \leq 200cm\} \times \{0 \leq y \leq 200cm\}$ for the plane $z = 0$, with the source being emitted in this plane at $t=0$. The delta function is represented by a square pulse. The snapshots are taken at times 0.1, 1.1, 2.1 and 3.1 ns and show the ballistic wave front with a diffusion component inside the circular region.}
    \label{fig:unbounded_propagation}
\end{figure}

\vspace{-2mm} The behavior of the solution can be seen in Figure \ref{fig:unbounded_propagation}, which shows the photon density inside the detector for several instants in a plane slice. 

\begin{figure}[!ht]
    \centering
\begin{tikzpicture}
    \draw (-1.3, 0) node[inner sep=0] {\includegraphics[trim = 50 23 0 0, clip, width=65mm]{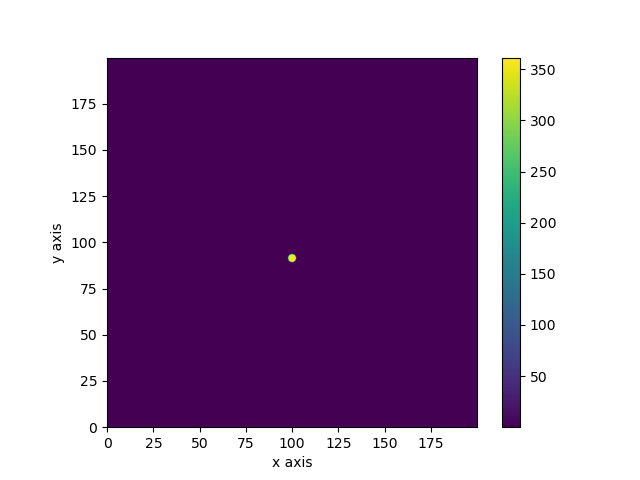}};
    \draw (-1.9, -2.9) node {x [cm]};
    \draw (-4.75, -0.5) node {\rotatebox{90}{y [cm]}};
    \draw (1.2, -0.5) node {\rotatebox{90}{photons / cm$^3$}};
    \draw (6.2, 0) node[inner sep=0] {\includegraphics[trim = 50 23 0 0, clip, width=65mm]{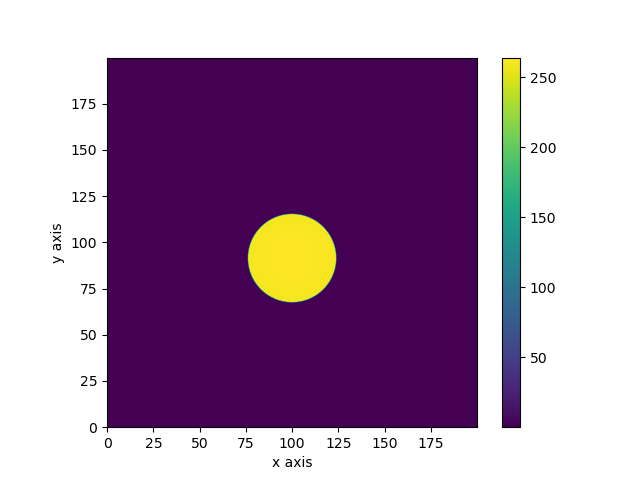}};
    \draw (5.6, -2.9) node {x [cm]};
    \draw (2.7, -0.5) node {\rotatebox{90}{y [cm]}};
    \draw (8.9, -0.5) node {\rotatebox{90}{photons / cm$^3$}};
    \draw (-1.3, -6) node[inner sep=0] {\includegraphics[trim = 50 23 0 0, clip, width=65mm]{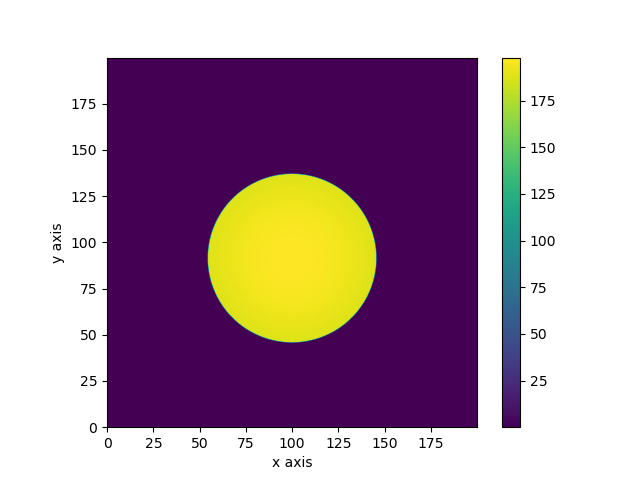}};
    \draw (-1.9, -8.9) node {x [cm]};
    \draw (-4.75, -6.5) node {\rotatebox{90}{y [cm]}};
    \draw (1.2, -6.5) node {\rotatebox{90}{photons / cm$^3$}};
    \draw (6.2, -6) node[inner sep=0] {\includegraphics[trim = 50 23 0 0, clip, width=65mm]{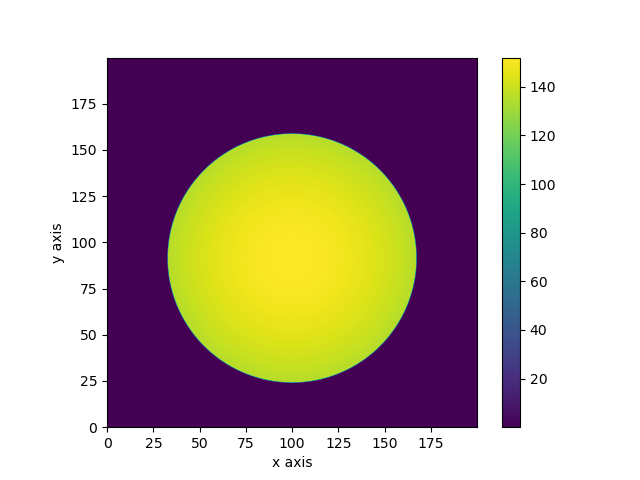}};
    \draw (5.6, -8.9) node {x [cm]};
    \draw (2.7, -6.5) node {\rotatebox{90}{y [cm]}};
    \draw (8.9, -6.5) node {\rotatebox{90}{photons / cm$^3$}};
\end{tikzpicture}
    \caption{The different figures illustrate just the diffusive component of the photon density in the square $ \{0 \leq x \leq 200cm\} \times \{0 \leq y \leq 200cm\}$ for the plane $z = 0$, with the source being emitted in this plane at $t=0$. The snapshots are taken at times 0.1, 1.1, 2.1 and 3.1 ns and show a homogeneous distribution at the first instants, but a gradient in the photon density at 3.1 ns}
    \label{fig:unbounded_propagation_diff}
\end{figure}
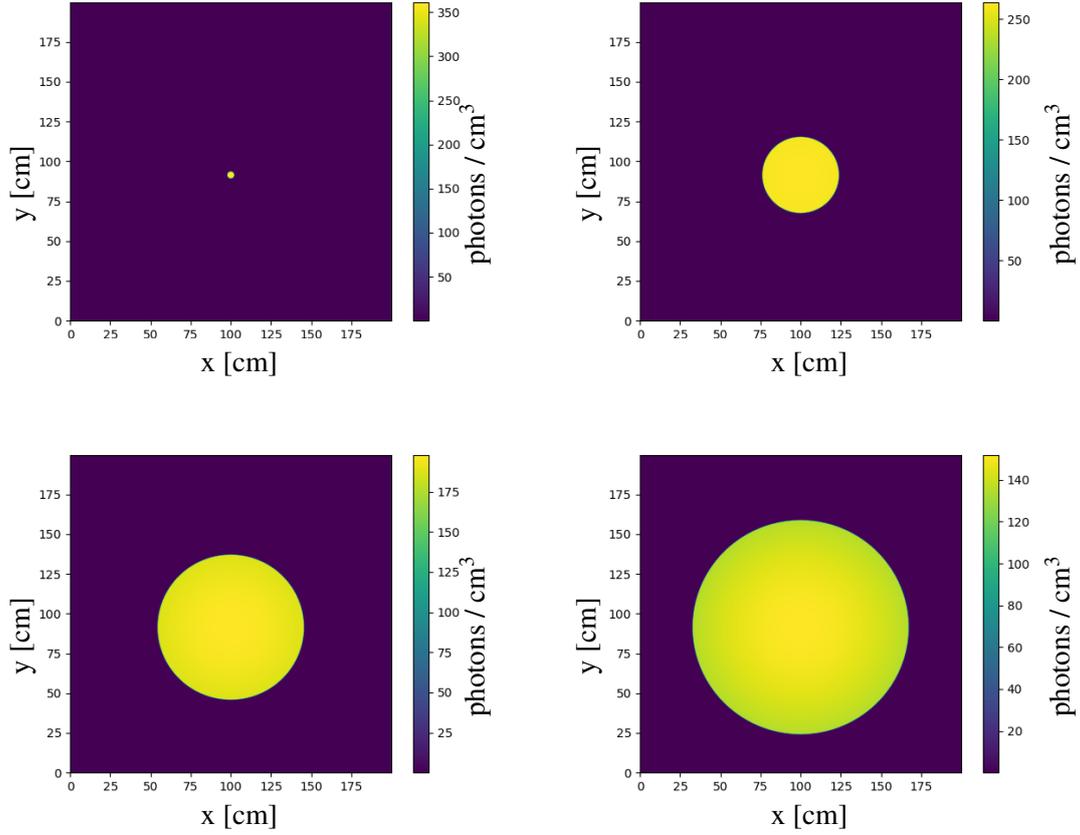

Figure \ref{fig:unbounded_propagation_diff} shows the diffusive component of the photon density, $\phi_{dif}$, inside the detector for several instants in a plane slice. 

Research is still needed to determine the correct way to consider boundary effects and photon flux through walls and detectors.

\section{Semi-analytical approach}

Garcia-Gamez, Green, and Szelc present a semi-analytical model to predict the amount of argon scintillation light observed by a light detector with a precision better than 10\%, based on the relative positions between the emission point and the detector in \cite{Szelc}.

They calculate the number of photons incident on a Photon Detector (PD) based on geometrical considerations as

$$
N_\Omega = e^{-\frac{d}{\lambda_{abs}}} \Delta E \cdot S_\gamma(\mathcal{E}) \frac{\Omega}{4\pi};
$$
where $d$ is the distance from the emission point to the PD, $\Delta E$ is the energy deposit, $S_\gamma(\mathcal{E})$ is the number of photons released per unit of deposited energy in a given electric field, and $\Omega$ is the solid angle of the PD from the emission point.

Comparing their predictions with Geant4 simulations, they find differences that can be fitted using Gaisser-Hillas (GH) functions:
$$
GH(d) = N_{max} \left( \frac{d-d_0}{d_{max}-d_0}\right)^{\frac{d_{max}-d_0}{\Lambda}}e^{\frac{d_{max}-d_0}{\Lambda}},
$$

 It is necessary then to find the best fit for the constants $(\Lambda, N_{max}, d_0, d_{max})$ in the GH functions, which demands some Geant4 simulations to obtain the corrections to the number
of detected photon. Thus making it a "semi-analytical" approach.

\begin{figure}[ht!]
    \centering
    \includegraphics[width=0.5\linewidth]{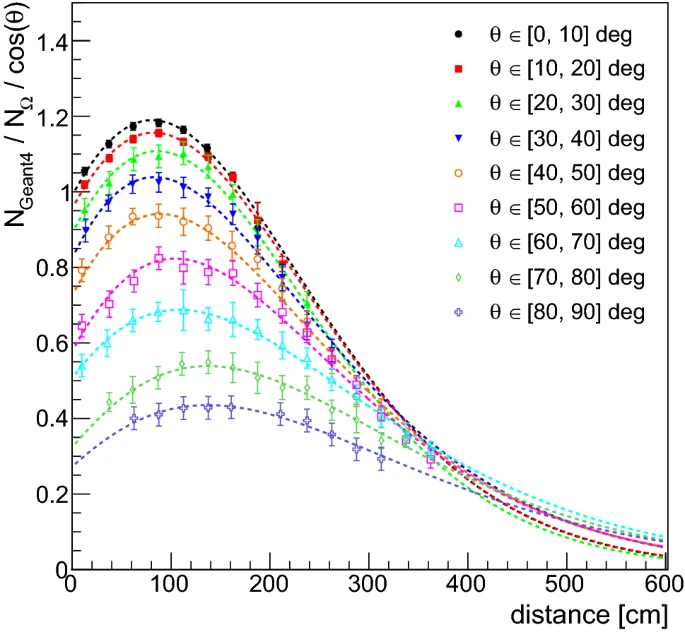}
    \caption{A figure from \cite{Szelc}, showing the ratio of the number of photons reaching a PD in Geant4 simulations and the number predicted by $N_\Omega$ divided by the cosine of the angle of the PD's normal to the line connecting the center of the detector to the emission point. The x-axis shows the distance from the emission point to the detector in cm.}
    \label{fig:eSzelc1}
\end{figure}

\section{Conclusion}

Photon propagation is an area important in diverse applications, from climate models and medical imaging to astrophysics and particle physics. There are diverse methods to simulate and predict the behavior of light, which are adequate for some situations, often requiring specific methods to match the real-world data of each case.

Differential equation methods that take into account relativistic effects and
attempt to account for the physical nature of light are promising. However, more research is needed to produce methods that work in diverse situations and are viable to calculate.

The semi-analytical approach of \cite{Szelc} seems to be a viable solution for LArTPCs, presenting a good approximation in most regions of the LArTPC and being faster than full Geant4 simulations on a large detector.

\acknowledgments

The studies that led to this work were conducted with the support of the National Council for Scientific and Technological Development of Brazil (CNPq).


\bibliographystyle{JHEP}
\bibliography{main.bib}

\end{document}